\documentclass[11pt,twoside]{article}
\usepackage{asp2010}
\usepackage{graphicx}
\usepackage{url}
\usepackage{epsfig}

\resetcounters

\begin{document}

\title{Spectropolarimetry of Chromospheric Magnetic and Velocity Structure Above Active Regions}
\author{T.A.~Schad$^{1,3}$, S.A.~Jaeggli$^2$, H.~Lin$^2$ and M.J.~Penn$^1$
\affil{$^1$National Solar Observatory, Tucson, AZ 85719, USA}
\affil{$^2$Institute for Astronomy, University of Hawai'i, Pukalani, HI 96768, USA}
\affil{$^3$Lunar and Planetary Laboratory, University of Arizona, Tucson, AZ 85721, USA}}

\begin{abstract}
Active regions often host large-scale gas flows in the chromosphere presumably directed along curved magnetic field lines. Spectropolarimetric observations of these flows are critical to understanding the nature and evolution of their anchoring magnetic structure.  We discuss recent work with the Facility Infrared Spectropolarimeter (FIRS) located at the Dunn Solar Telescope in New Mexico to achieve high-resolution imaging-spectropolarimetry of the Fe I lines at 630 nm, the Si I line at 1082.7 nm, and the He I triplet at 1083 nm.  We present maps of the photospheric and chromospheric magnetic field vector above a sunspot as well as discuss characteristics of surrounding chromospheric flow structures.
\end{abstract}

\section{Introduction}

The He I 10830~\AA~triplet has been known for some time to be a promising spectral diagnostic for both Doppler shifts and magnetically-sensitive polarization signatures formed in the upper chromosphere, and in particular above active regions \citep{lites85,penn02,harvey71}.  Recently, our understanding of the mechanisms inducing or modifying its polarization has greatly improved \citep[see review by][]{lagg07a}, which along with advances in instrumentation, now allow reliable inference of the full chromospheric magnetic field vector.

In this article we discuss work with the recently commissioned Facility Infrared Spectropolarimeter (FIRS) to acheive spectropolarimetric observations in both the photosphere and chromosphere over a large field-of-view at high spatial resolution.  We examine the height-dependence of the full magnetic field vector above a sunspot as well as describe the polarization signatures of the diversity of chromospheric structures hosting gas flows.

\section{Data Description}

\subsection{The Facility Infrared Spectropolarimeter (FIRS)} 
\label{sect:obs_firs}

\begin{figure}[!Ht]
\begin{center}
\includegraphics[width=0.99\textwidth, height=0.83\textheight]{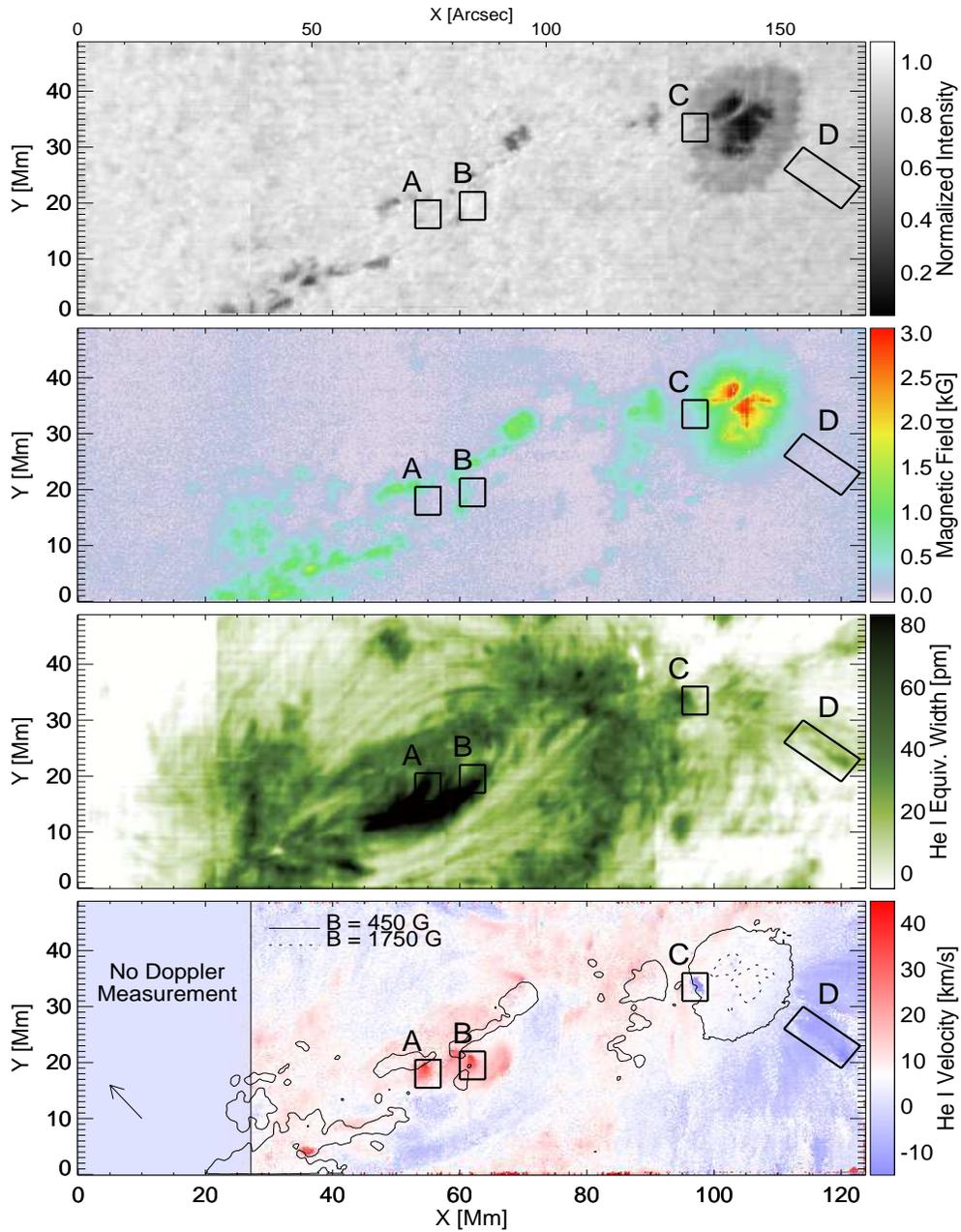}
\end{center}
\caption{FIRS observations of NOAA active region 11072 (36\deg W, 24\deg S) on 7 July 2009 beginning at 12:27:58 UT.  Solar north is up. Top two figures give the continuum intensity and photospheric magnetic field inferred from the 630.25 nm Fe I line. Contours of the photospheric field are provided on the He I velocity map. The arrow denotes the direction of disk center.}
\label{fig:firs_maps}
\end{figure}

The Facility Infrared Spectropolarimeter \citep[FIRS:][]{jaeggli08} located at the 76 cm aperture Dunn Solar Telescope (DST, Sacramento Peak, New Mexico), consists of an off-axis reflecting, Littrow-configuration spectrograph designed to perform simultaneous dual-beam spectropolarimetry at visible and infrared wavelengths.  For fast scanning of large solar regions, FIRS uses narrowband filters in a multi-slit mode to image spectra from multiple parallel slits on the same detector.  Diffraction-limited observations can be acheived with concurrent use of the telescope's High Order Adaptive Optics system \citep[HOAO:][]{rimmele04}.

On 2009 July 7 we made observations of NOAA AR 11024\,--\,located at 36\deg W, 24\deg S ($\mu=0.71$)\,--\, using FIRS in its wide-field f/36 configuration and with the aid of the HOAO system.  By stepping the solar image across a four-slit unit in steps of $0.29''$, two full areas scans with a $168''\times 70''$ field of view were obtained (see Fig.~\ref{fig:firs_maps}).  At each scanning position, the full Stokes vector was measured by both the visible and infrared instrument arms using a four-state efficiency balanced modulation sequence.  The visible arm measured the photospheric Fe I spectral lines at 630.15 and 630.25 nm with a bandpass from 630.09 to 630.30 nm, spectral dispersion of 1 pm, and a spatial sampling along the slit of $0.08''pix^{-1}$.  Meanwhile the infrared arm measured the chromospheric He I triplet at 1083 nm \citep[see][for line parameters]{lagg07b}.  FIRS is designed also to measure the nearby photospheric Si I line at 1082.7 nm within the filter bandpass for each slit; however, the instrument setup on 2009 July 7 imaged this line only in the slit scanning across the sunspot.  Also, only a portion of the He I triplet was measured in the eastern-most slit.  The spectral dispersion of these infrared measurements is 3.65 pm with a spatial sampling along the slit of $0.15''pix^{-1}$.  Each full area scan took 22 minutes to complete with an acquired polarimetric RMS noise level for the infrared measurements of $\sim2\times 10^{-3}$ I$_{\rm C}$ at 0.29$''$ spatial resolution.

\subsection{Calibration and Inversion Methods}

The usual methods for flat-field and dark current correction were applied to our FIRS observations.  Telluric line positions were used to correct for slit curvature and spectral drift during scanning.  Disk-center line positions were used for a wavelength calibration.  Two separate methods for polarimetric calibration were used.  Using a linear polarizer and a quarter-wave plate inserted at the telescopic exit port upstream of the instrument, we determine the instrumental Mueller matrix for both the visible and infrared FIRS arms.\footnote[1]{For more information see \url{http://www.ifa.hawaii.edu/~jaeggli/wip/polcal/pol_document.pdf}}  After removing the instrumental polarization crosstalk, we apply the polarization calibration method of \citet{kuhn94} using the Stokes profiles of the strong fields of the outer umbra/inner penumbra region of the sunspot measured with the photospheric spectral lines.

We invert our Stokes spectra using HeLIx$^{+}$, the He-Line Information Extractor$^{+}$, a flexible inversion code described by \citet{lagg04, lagg07b}. A one magnetic component Milne-Eddington (ME) model is implemented for inferring the photospheric magnetic vector using the Fe I 630.25 nm spectral line (see Figure~\ref{fig:firs_maps}).  We perform a similar inversion for the photospheric Si I 1082.7 nm line (see Figure~\ref{fig:mag_vec}).  Our inversion of the He I triplet at 1083 nm required the use of a multi-component model.  In nearly all spectra, a Helium component of relatively small Doppler shift is evident.  In locations of strong supersonic downflows (e.g. locations `A' and `B' in Figure~\ref{fig:firs_maps}), two well-separated atmospheric components are recognizable in the intensity spectra.  We thus use a two component ME model with one component restricted to slow Doppler velocities; the other component is allowed wide velocity bounds.  In all maps within this article, only the results of the fast component are displayed.  As the RMS noise in this data set is too high to address small-scale magnetic fields, we only take into account the Zeeman and Paschen-Back effects, which give reliable inversions for field strengths larger than $\sim1000$ G \citep{centeno09}.  We also weight the contibutions of the Stokes profiles to optimize the fits to Stokes I and V.  In future FIRS measurements with improved signal-to-noise, the full line calculation incorporating atomic level polarization and the Hanle effect will be necessary.  An inversion code named HAZLE solves for each of these effects in a cloud-type atmospheric model \citep{asensio_ramos08} and is also available within the HeLIx$^{+}$ architecture.

\section{Multi-Height Sunspot Magnetic Structure}

\begin{figure}[!Ht]
\begin{center}
\includegraphics[width=0.99\textwidth]
{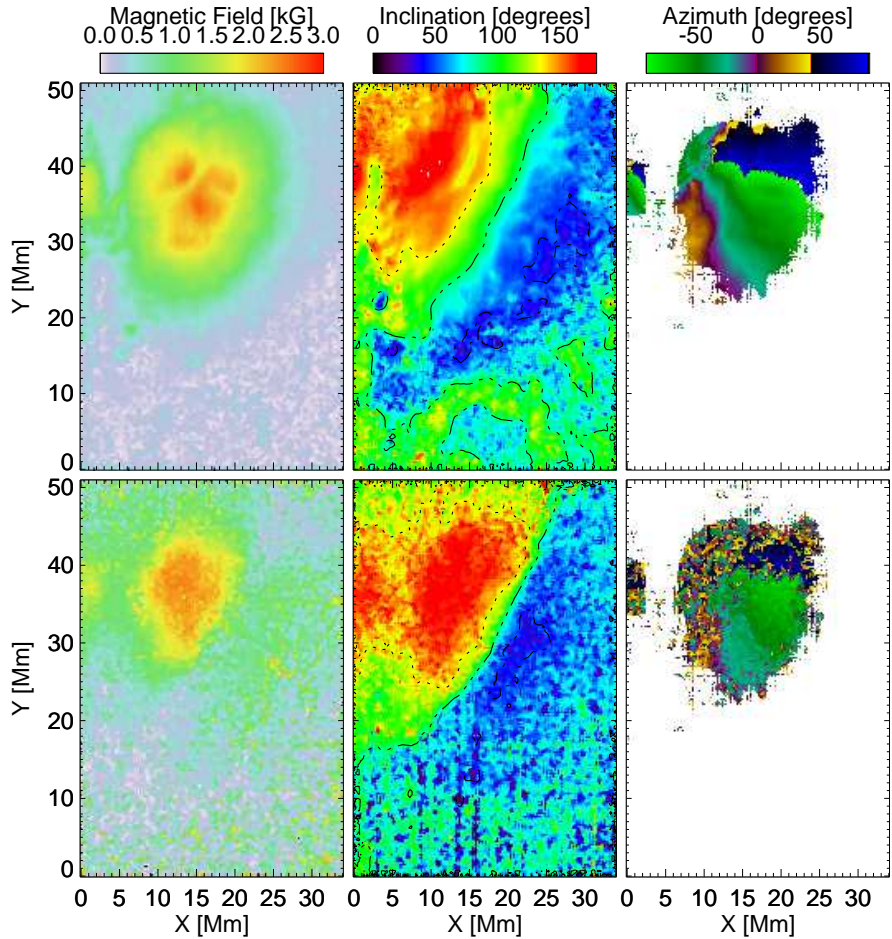}
\end{center}
\caption{(top) Magnetic Field strength, inclination, and azimuth (in the geometry of the observer; azimuth zero point is directed westward) derived from the Si I 1082.7 nm spectral line in the western-most FIRS slit. (bottom) The chromospheric field vector resultant from an analysis of the He I 1083 nm triplet.  Azimuth values are shown only for those locations meeting a fitness criterion.  Inclination contours are given for 45$\deg$, 90$\deg$, and 135$\deg$.}
\label{fig:mag_vec}
\end{figure}

The full magnetic field vectors derived from the photospheric Si I 1082.7 nm line and the chromospheric He I 1083 nm triplet within the large sunspot of AR 11024 are presented for comparison in Figure~\ref{fig:mag_vec}.  Inclination and azimuth values are given in the observer's reference frame since transforming to the solar frame requires knowledge of the height of formation for the He I triplet, which is not precisely known.  The determined field vector in the chromosphere displays remarkable resemblance to that inferred at the photospheric level, and the directional variation of the azimuth measured at the two heights shows a near direct correspondence.  Although not shown, the field geometry inferred from the Fe I 630.2 nm line, which is formed within 100 km of the Si I 1082.7 nm line, displays nearly identical features to that inferred from the Si I line.  Despite the steep viewing angle ($\mu=0.71$), the magnetic field in the chromosphere shows similar spatial structure as well as a decrease in strength as compared to the photosphere.  Along the line-of-sight, the magnetic field above dark umbral features decreases by 0.0-0.6 G km$^{-1}$, if one assumes a 2000 km height difference between Si I and the He I triplet.  A simple radial dependence (with respect to sunspot center) of this decrease cannot be determined due to the complications invoked by the light bridge, which as expected does not exhibit a chromospheric magnetic field signature.  The inferred inclination angle also compares favorably between the photosphere and chromosphere.  At each height the inclination map reveals a field `turn-over' where the field changes from being directed towards the observer (inclination $< 90$) to away (inclination $> 90$), which is due to both the viewing angle and the expansion of the magnetic field.  Iso-curves of inclination shown in Figure~\ref{fig:mag_vec} delineate the horizontal variation of the inclination angle at the two levels.  While the inclination angle inferred from the Si I and Fe I lines are nearly identical, the chromospheric inclination angle changes much more rapidly in the direction away from disk center suggesting a large expansion of the field at this height.

\section{Chromospheric Flows}

Considering the promise of the He I triplet as a velocity diagnostic, the ideal application for He I polarimetry studies is for strong flows in the chromosphere.  Supersonic downflows in the chromosphere have been studied with He I spectropolarimetry and shown to be both commonplace in active regions and suggestive of chromospheric fine structure \citep{aznar_cuadrado05, lagg07b}.  We too observe such supersonic downflows up to 40 km sec$^{-1}$ in AR 11024 (see locations marked `A' and `B' within Figure~\ref{fig:firs_maps}); although, here we discuss two different chromospheric flow features.

\subsection{Evidence for a Strong Chromospheric Upflow}

\begin{figure}[!Ht]
\begin{center}
\includegraphics[width=0.95\textwidth]{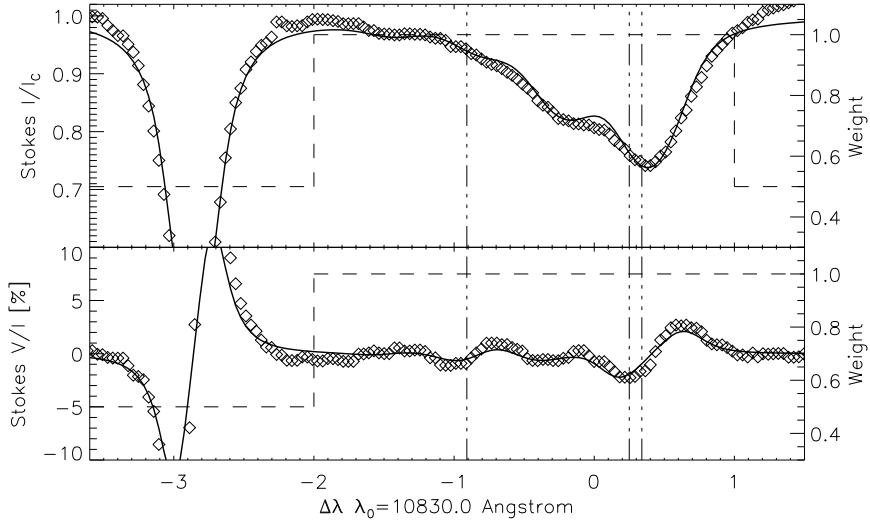} 
\end{center}
\caption{Observation of a strong chromospheric upflow:  Stokes I and V/I spectra (diamonds) are shown for the greatest blue-shifted pixel inside box `C' shown in Figure~\ref{fig:firs_maps}.  HeLIx$^{+}$ fits (solid lines) for the line-of-sight magnetic field component in a Milne-Eddington atmosphere with a fast and slow flow-component give the following values:  $B_{slow}=551$ G, $V_{LOS,slow}=2434$ m sec$^{-1}$; $B_{fast}=268$ G, $V_{LOS,fast}=-14605$ m sec$^{-1}$.  Vertical dot-dashed lines give the rest positions for the He I triplet lines.  The dashed lines gives the wavelength-dependent weighting value considered within the HeLIx$^{+}$ fit.}
\label{fig:blue_shift}
\end{figure}

A small feature exhibiting a strong chromospheric blue-shift exists on the disk-ward side of the large umbra in Figure~\ref{fig:firs_maps} (labeled `C').  While numerous observations of prominent downflows and strong, presumably-horizontal flows (i.e. inverse Evershed effect, see Section~\ref{sec:inv_ever}) have been discussed by previous authors, little mention has been made of strong upflows.  The feature observed here is recognized in both acquired FIRS maps (22 minute difference) and is located above a portion of the large sunspot where the penumbra is reduced in extent.  The feature exhibits very limited spatial extent ($<2''$) and is therefore interpreted to be a flow directed upwards.  The intensity spectrum, shown in Figure~\ref{fig:blue_shift}, reveals the presence of two components\,--\,a nearly-unshifted component and a strongly blue-shifted component.  The Stokes V/I profile, also shown in the figure, supports the presence of two different magnetic components.  Linear polarization measurements are not shown since the noise is too high.  We invert this spectra using HeLIx$^{+}$ to constrain the line-of-sight magnetic field strength for the two atmospheric components.  The contribution of Q and U to the fit is ignored.  Repetitive fits using a large number of iterations within the HeLIx$^{+}$ PIKAIA genetic algorithm give the statistically-best possible fit (values given in Figure~\ref{fig:blue_shift}).  The strongly blueshifted component has a weaker line-of-sight magnetic field strength indicating a change in field direction and/or height of the observed atmospheric component within the solar atmosphere.  Linear polarization measurements with higher sensitivity would be advantageous.

\subsection{The Chromospheric Inverse Evershed Effect}\label{sec:inv_ever}

Complementary to the outward-directed (relative to sunspot center) photospheric Evershed flow, an inward-directed flow, dubbed the inverse Evershed effect, exists in laterally-extended superpenumbral fibrils in the chromosphere.  The source of this flow is not well understood \citep[see][and references therein]{alissandrakis88}, as well as its possible influence on active region twist \citep{bala04}.  Advancing our ability to probe the magnetic field in these features through spectropolarimetry may yield a more accurate determination of the velocity vector along individual fibrils as well as their magnetic connection to the photosphere and/or corona. The inverse Evershed effect is a prominent feature on the southwestern side of AR 11024 in Figure~\ref{fig:firs_maps}.  Individual flow channels are recognizable in these high-resolution observations and match features in the equivalent width maps for He I.  Line-of-sight velocity values peak between 10 and 12 km sec$^{-1}$ within the channels indicated by box 'D'.  While these values are larger than those found by \citet{penn02}, they are consistent with the multi-height measurements of the inverse Evershed flow discussed by \citet{alissandrakis88} as the flow velocities derived here are lower than those seen in the lower corona.

The polarimetric signals within these features are considerably hard to measure.  We do not give any derivation of the field geometry of these features since the noise in QU for these data is too large.  Only in the strong absorption feature within box 'D' does the total polarization ($\sqrt{Q^{2}+U^{2}+V^{2}}/I$) show values slightly higher than the noise ($\sim2\times 10^{-3}$ I$_{\rm C}$).  Additionally, considerable changes are seen between the two FIRS maps acquired with a time difference of 22 minutes.  Both these temporal changes and large horizontal gradients in the observed LOS flow velocity make these features a particularly difficult observable for He I triplet spectropolarimetry.  However, the appropriate targeting of these features with FIRS should be able to achieve higher polarimetric sensitivity.

\section{Concluding Remarks}\label{sec:conclusion}

Active region chromospheres exhibit a wide range of magnetic structures with strong gas flows.  Spectropolarimetric diagnostics of these phenomena are expanding and improving.  Observational advances, like the wide-field fasting-scanning FIRS, are necessary to study chromospheric features of large lateral extent.  Further multi-line studies of chromospheric flows are necessary.  Observational work employing both FIRS and the Interferometric BiDimensional Spectrometer \citep[IBIS:][]{cavallini06,reardon08}, which has been upgraded for full Stokes polarimetry, is in progress.

\acknowledgements We extend our thanks to the SPW 6 SOC for their support of graduate students\,--\,in particular Sarah Jaeggli and Tom Schad.  This work was supported by the University of Hawaii's Institute for Astronomy as well as the National Solar Observatory, operated by the Association of Universities for Research in Astronomy, Inc. (AURA), under cooperative agreement with the National Science Foundation.  The FIRS project was funded by the National Science Foundation Major Research Instrument program, grant number ATM-0421582. 

\bibliography{schad_spw6}

\end{document}